\documentclass[12pt,prd,draft,nofootinbib]{revtex4}

\usepackage{amsfonts}
\usepackage{amssymb}
\usepackage{amsmath}
\usepackage{bm}
\usepackage{latexsym}

\newcommand{\beq}{\begin{eqnarray}}
\newcommand{\eeq}{\end{eqnarray}}
\newcommand{\be}{\begin{equation}}
\newcommand{\ee}{\end{equation}}
\def\la{\mathrel{\mathpalette\fun <}}

\def\fun#1#2{\lower3.6pt\vbox{\baselineskip0pt\lineskip.9pt
\ialign{$\mathsurround=0pt#1\hfil ##\hfil$\crcr#2\crcr\sim\crcr}}}

\newcommand{{\SD}}{\rm SD}

\newcommand{\vep}{\bm p}
\newcommand{\veq}{\mbox{\boldmath${\rm q}$}}

\newcommand{\lan}{\langle}
\newcommand{\ran}{\rangle}

\begin{document}

\title{ The ratio of decay widths of  $X(3872)$ to  $ \psi^{\prime}\gamma $  and $ J/\psi\gamma$
as a test of the  $X(3872)$  dynamical structure}

\author{\firstname{A.M.}~\surname{Badalian}}
\email{badalian@itep.ru}

\author{\firstname{V.D.}~\surname{Orlovsky}}
\email{orlovskii@itep.ru}

\author{\firstname{Yu.A.}~\surname{Simonov}}
\email{simonov@itep.ru} \affiliation{Institute of Theoretical and
Experimental Physics, Moscow, Russia}

\author{\firstname{B.L.G.}~\surname{Bakker}}
\email{b.l.g.bakker@vu.nl} \affiliation{Department of Physics and Astronomy,
Vrije Universiteit, Amsterdam, The Netherlands}

%\author{A.M.Badalian, V.Orlovsky, Yu.A.Simonov, and B.L.Bakker^a}
%\affiliation{State Research Center, Institute of Theoretical and
%Experimental Physics, Moscow, 117218 Russia}
%\email[E-mail;]{badalian@itep.ru}
%
\date{\today}

\begin{abstract}
Radiative decays of $X(3872)$ with $J^{PC}=1^{++}$ are studied in
the coupled-channel approach, where the $c\bar c$ states are
described by relativistic string Hamiltonian, while for the decay
channels $DD^*$ a string breaking mechanism is used. Within this
method a sharp peak and correct mass shift of the $2\,{}^3P_1$
charmonium state just to the $D^0D^{*0}$ threshold was already
obtained for a prescribed channel coupling to the $DD^*$ decay
channels. For the same value of coupling the normalized wave
function (w.f.) of $X(3872)$ acquires admixture of the
$1\,{}^3P_1$ component with the weight
$c_1=0.153~(\theta=8.8^\circ$), which increases the transition
rate $\Gamma(X(3872)\rightarrow J/\psi\gamma)$ up to 50-70~keV,
making the ratio $R=\frac{\mathcal{B}(X(3872)\rightarrow
\psi^{\prime}\gamma)}{\mathcal{B}(X(3872)\rightarrow J/\psi
\gamma)}=0.8\pm 0.2~(th)$ significantly smaller, as compared to
$R\simeq 5$ for $X(3872)$ as a purely $2\,{}^3P_1$ state.
\end{abstract}

\maketitle

\section{Introduction}
The $X(3872)$ was discovered by Belle as a narrow peak in $J/\psi \pi\pi$
invariant mass distribution in decays $B\rightarrow J/\psi\pi\pi K$
\cite{ref.1} and later confirmed by the CDF, D0, and BaBar Collaborations
\cite{ref.2}. It has several exotic properties, very small width $\Gamma<
2.3$~MeV and the mass very close to the $D^0D^{*0}$ threshold
\cite{ref.3},\cite{ref.4}. The even charge parity $C=+$ of  $X(3872)$ is now
well established \cite{ref.5}, while two most plausible assignments for its
quantum numbers, $J^{PC}=1^{++}$ and $2^{-+}$, are still discussed
\cite{ref.6}, \cite{ref.7}.

To understand the nature of $X(3872)$ a special role belongs to radiative
decays, $X(3872)\rightarrow J/\psi \gamma$ and $X(3872)\rightarrow
\psi^{\prime}\gamma$. The first evidence for the decay $X(3872)\rightarrow
J/\psi \gamma$ was obtained by Belle \cite{ref.8} and confirmed by BaBar
\cite{ref.9}; later the BaBar has observed radiative decay $X(3872)\rightarrow
\psi^{\prime}\gamma$ with the branching fraction ratio
$R=\frac{\mathcal{B}(X(3872)\rightarrow \psi^{\prime}
\gamma)}{\mathcal{B}(X(3872)\rightarrow J/\psi\gamma)}=3.4\pm 1.4$
\cite{ref.10}. Knowledge of this ratio is of special importance for theory,
because the rates of these radiative decays vary widely in different
theoretical models \cite{ref.11}-\cite{ref.15}. If $X(3872)$ is considered as a
conventional $2\,{}^3P_1$ charmonium state, then the characteristic value of
$\rm R$ is rather large, $R\simeq 4-6$ \cite{ref.12}- \cite{ref.14}, being in
general in agreement with the BaBar number $3.4\pm 1.4$. In molecular picture
the radiative decay $X(3872)\rightarrow \psi^{\prime}\gamma$ is suppressed and
the ratio $R$ should be much smaller \cite{ref.11}.

 However, in 2010 on a
larger sample of decays $\rm B\rightarrow X(3872)K)$ the Belle has not found
evidence for the radiative decay $X(3872)\rightarrow \psi^{\prime}\gamma$,
giving the upper limit  $\rm R< 2.1$ \cite{ref.16}. This number does not agree
with the representation of $X(3872)$ as a purely $2\,{}^3P_1$ charmonium state.

Existing experimental uncertainty calls for new studies of coupled-channel (CC)
effects for $X(3872)$. In \cite{ref.13} the authors have anticipated ``... a
significant $DD^*$ component in $X(3872)$, even if it is dominantly a $c\bar c$
state". For that the $\,{}^3P_0$ model was used in \cite{ref.13} and the
Cornell many-channel model was considered in \cite{ref.14}. However, in spite
of these many-channel calculations there predicted values of $R$ have appeared
to be close to those obtained for $X(3872)$ as a purely $2\,{}^3P_1$ charmonium
state: $R=5.8$ in \cite{ref.13} and $R=5.0$ in \cite{ref.14}.

Here we consider $X(3872)$ with $J^{PC}=1^{++}$ in the CC approach, where a
coupling to the $DD^*$ channels is defined by the relativistic string-breaking
mechanism, which was already applied to $X(3872)$ in \cite{ref.17},
\cite{ref.18}, explaining it as $2\,{}^3P_1$ charmonium state shifted down and
appearing as a sharp peak just at the $D^0D^{*0}$ threshold. Besides, the
scattering amplitude and a production cross section were calculated there,
being in qualitative agreement with experiment. Here we apply this method for
calculations of the radiative decay rates for $X(3872)$ and show that due to
the same CC mechanism (with the coupling of the same strength) an admixture of
the $1\,{}^3P_1$ component to the $X(3872)$ w.f. appears to be not large, $\sim
15\%$; nevertheless, this component strongly affects the value of the partial
width $\Gamma_1=\Gamma(X(3872)\rightarrow J/\psi \gamma)$ and decreases the
ratio $\rm R$.

\maketitle

\section{Coupled-channel mechanism}

We use here the string decay Lagrangian of the ${}^3P_0$ type for
the decay $c\bar c\rightarrow (c\bar q)(\bar c q)$ \cite{ref.18}:

\begin{equation}
\mathcal{L}_{sd} =\int \bar \psi_q M_\omega \psi_q
~d^4x\label{eq.1}
\end{equation}
where the light quark bispinors are treated in the limit of large $m_c$, which
allows us to go over to the reduced $(2\times 2)$ form of the decay matrix
elements (m.e.). Also to simplify calculations the actual w.f. of $c\bar c$
states, calculated in \cite{ref.19} with the use of the relativistic string
Hamiltonian (RSH), is fitted here by five (or three) oscillator w.f. (SHO),
while the $D$ meson w.f. is described by a  single SHO term  with few percent
accuracy  with the parameter $\beta\simeq 0.48$. In this case the factor
$M_{\omega}$ in (\ref{eq.1}) is $M_{\omega}\simeq \frac{2\sigma}{\beta}\simeq
0.8$~GeV, which produces correct total width of $\psi(3770)$ and it will be
used below.

 The transition m.e. for the decays $(c\bar c)_n\rightarrow (D\bar
D), (DD^*), (D^*D^*)$ are denoted here as $n\rightarrow n_2,n_3$, and in the
$2\times 2$ formalism this m.e. reduces to

\be
  J_{nn_2n_3} (\vep) =\frac{\gamma}{\sqrt{N_c}} \int \bar{
y_{123}} \frac{d^3\veq}{(2\pi)^3} \Psi^+_n (\vep +\veq) \psi_{n_2} (\veq)
\psi_{n_3} (\veq).\label{eq.2}\ee Here $\gamma=\frac{2M_{\omega}}{<m_q+U -V_{D}
+\epsilon_0>}$, where average of the Dirac denominator (with scalar confining
potential $U=\sigma r$ and vector potential $V_D=-\frac{4\alpha}{3 r}$) is
calculated and yields $\gamma=1.4$. The factor $\bar y_{123}$ contains a trace
of spin-angular variables (for details see \cite{ref.18}, \cite{ref.20}).

 The
intermediate decay channel, like $DD^*$, induces an additional interaction
``potential" $V_{CC}(\vec{q},\vec{q'},E)$ (here the quotation marks imply
nonlocality and energy dependence of this potential):

\be V_{CC} (\veq, \veq', E) = \sum_{n_2n_3} \int \frac{d^3\vep}{(2\pi)^3}
\frac{X_{n_2n_3}(\veq, \vep) X_{n_2n_3}^+ (\veq', \vep)}{E-E_{n_2n_3} (\vep)},
\label{eq.3}\ee where

\be X_{n_2 n_3} (\veq, \vep)=\frac{\gamma}{\sqrt{N_c}} \bar y_{123} (\veq,
\vep) \psi_{n_2} (\veq-\vep) \psi_{n_3} (\veq-\vep).\label{eq.4}\ee Using
(\ref{eq.3}) and (\ref{eq.4}) one can find how the energy eigenvalues (e.v.)
and the w.f. of a state $(c\bar c)_n)$ change due to the interaction $V_{CC}$.
In particular, in the first order of perturbation theory one has

   \be E^{(1)}_n = E_n + w_{nn} (E_n),\label{eq.5}\ee

  \be \psi_n^{(1)} = \psi_n + \sum_{m\neq n} \frac{w_{nm} (E_n)}{E_n-E_m}
  \psi_m, \label{eq.6}\ee
where $\psi_n,~E_n$ refer to the unperturbed $(c\bar c)_n$ system
and  the m.e. $w_{nm}(E)$ is

 \be w_{nm} (E) = \int \psi_n (\veq ) V_{CC} (\veq, \veq', E) \psi_m (\veq')
 \frac{d^3\veq}{(2\pi)^3}\frac{d^3\veq'}{(2\pi)^3}=\label{eq.7}\ee
 $$= \int \frac{d^3\vep}{(2\pi)^3}\sum_{n_2n_3} \frac{J_{nn_2n_3}(\vep)
 J^+_{mn_2n_3} (\vep)}{E-E_{n_2m_3}(\vep)}.$$

Notice, that if the CC interaction is strong, then one should take
into account this interaction to all orders, summing the infinite
series over $V_{CC}$ (or $w_{nm}$). As a result, one obtains the
full Green's function for an arbitrary $Q\bar Q$ system (our
result formally coincide with those from \cite{ref.21}, although
differences occur in concrete expressions for $w_{nm}$, because
our interaction and decay mechanism differ from those in
\cite{ref.21}):

   \be G_{Q\bar Q }(1,2;E) = \sum_{n,m} \Psi^{(n)}_{Q\bar Q} (1) (\hat E - E + \hat
   w)^{-1}_{nm} \Psi^{+(m)}_{Q\bar Q} (2),\label{eq.8}\ee
here $\hat E_{nm}=E_n \delta_{nm}$. Then the energy e.v. are to be
found from the zeros of the determinant

    \be \det (E-\hat E-\hat w) =0.\label{eq.9}\ee
If the mixing of the state $n$ with other states, $m\neq n$, is
neglected, one obtains a nonlinear equation for the e.v. $E_n^*$
with a correction:

\be E_n^{(*)} = E_n + w_{nn} (E^{(*)}_n). \label{eq.10}\ee

 Note that $E_n^*$
can be a complex number and occur on the second sheet in the complex plane with
a cut from the threshold $n_2n_3$ to infinity. The expression (\ref{eq.10}) was
used in \cite{ref.17} to find the position of the pole for shifted $2\,{}^3P_1$
charmonium state, while its unperturbed value $E_2$ was calculated with the use
of RSH \cite{ref.22} to be $E_2=3948\pm 10$~MeV. In \cite{ref.17} the decays
have included $D_0D_0^*$ and $D^+D^-$ channels and $\gamma$ was used as a free
parameter. Then the position of the resulting pole (\ref{eq.10}) and the
production cross section have been calculated, giving the pole position exactly
at the $D_0D_0^*$ threshold for $\gamma=1.2$, which is close to expected number
$\gamma=\frac{M_\omega}{\lan m_q+u-V_D+\varepsilon_0\ran } = 1.4$ for $M_\omega
=0.8$ GeV. (From Fig.2 in \cite{ref.17} one can see that for this value of
$\gamma$ the production curve agrees well qualitatively with experiment.)

Knowing the mass shift of  $X(3872)$, one has also to find next order
corrections to the w.f., thus going beyond no-mixing approximation. At first,
we use perturbation theory, when admixture of the states $m\,{}^3P_1\equiv
m1^{++}$ to the $2\,{}^3P_1$ state is given by the (\ref{eq.6}),

\be c^{(1)}_m = \frac{w_{2m}}{E_2-E_m}, ~~ m=1,3,4,....\label{eq.11}\ee
Calculations give $w_{21}=0.085$~GeV, $w_{23}=-0.0008$~GeV, while
$E_2-E_1=0.425$~GeV, so that  main contribution comes from the $m=1$
($1\,{}^3P_1)$ state with the mixing parameter

\be c^{(1)}_1 = \frac{w_{21}}{E_2-E_1} \cong 0.20.\label{eq.12}\ee This
correction $c_1^{(1)}$ is not small and calls for a more accurate calculations,
beyond perturbation theory.

To this end we first write general expressions for the yield of particles
$\gamma, \pi, \rho,\omega$ etc. from the  system,  originally born as a $(Q\bar
Q)$ system  in  $e^+e^-$ or $B$ meson decay by  an operator $\hat B$. The
Green's function for the system can be written as \be G_{Q\bar
Q}^{(\mathcal{B}\mathcal{B})} =\sum_{n,m} (\hat B \psi_n) \left( \frac{1}{\hat
E - E + \hat w }\right)_{nm} (\hat B \psi_m), \label{eq.13}\ee where  e.g.
$(\hat B \psi_n)\sim \psi_n (0) $ for  the $e^+e^-$ production. We now can
attribute particle $(i)$ production from $Q,\bar Q$ lines adding the
corresponding self energy parts to $\hat E$, $(\hat E)_{nm} = E_n \delta_{nm} +
\sum^{(i)}_{nm} (E),$ and the production from light quark lines to $w_{mn}
(e)$, $w_{mn} (E) \to w_{mn} (E) + w_{mn}^{(i)} (E)$. Then the yield for
particles $i$ can be written as \be Y_i (E) = \sum_{n,m,l,q} (\hat B \psi_n)
\left( \frac{1}{\hat E - E + \hat w }\right)_{nm}  \frac{\Delta^{(i)}}{2i}
(\Sigma^{(i)}_{ml} + w^{(i)}_{ml} )\left( \frac{1}{\hat E - E + \hat w
}\right)_{lq}^*  (\hat B \psi_q)^*. \label{eq.14}\ee One can define now
effective $(Q\bar Q)$ wave function $\Psi_{Q\bar Q}$, which actually
participates at the vertex of emission of particles $i$, \be \Psi_{Q\bar
Q}^{(\mathcal{B})}=\sum_{k,l} \Psi_k \left( \frac{1}{\hat E - E + \hat w
}\right)_{kl} (\hat B\psi_l) \equiv \sum_k a_k \Psi_k.\label{eq.15}\ee

Keeping only two eigenfunctions, one has \be a_1 = \frac{E_2 -E + w_{22}}{\det}
(\hat B \psi_1) - \frac{w_{21}}{\det} (\hat B \psi_2)\label{eq.16}\ee \be a_2 =
\frac{E_1 -E + w_{11}}{\det} (\hat B \psi_2) - \frac{w_{12}}{\det} (\hat B
\psi_1),\label{eq.17a}\ee where $\det \equiv \det(\hat E - E+ \hat w)$.

>From (\ref{eq.17}) one obtains the ratio \be \frac{a_1}{a_2} = \frac{w_{21} +
(E_R-E) \frac{(\hat B\psi_1)}{(\hat B\psi_2)}}{\Delta_1 - w_{11} +
w_{12}\frac{(\hat B\psi_1)}{(\hat B\psi_2)}}, ~~ \Delta_1 \equiv E_R-E_1= 362~{
\rm MeV}.\label{eq.18a}\ee

Neglecting $\frac{(\hat B\psi_1)}{(\hat B\psi_2)},$  and for $E=E_R = 3872$ MeV
one obtains \be\frac{a_1}{a_2} = \frac{w_{21}}{\Delta_1 -w_{11}}
=0.179,\label{eq.19a}\ee while approximating in  $B$-decay production
$\frac{(\hat B\psi_1)}{(\hat B\psi_2)}\approx \frac{\psi'_1(0)}{\psi'_2(0)}
\approx 0.85$ one obtains \be \frac{a_1}{a_2} \cong 0.155. \label{eq.20a}\ee

One can see in Table I, that the $n=3 ~^3P_1$ state gives a negligible
admixture.
 The values of $w_{mn}$ for $E=E_R$ are computed according to Eq. 7 with w.f.
 obtained in \cite{ref.19} and are given in Table I.

\begin{table}[h]
\caption{The m.e. $w_{nm}$ (in GeV) between $n\,{}^3P_1$ and $m\,{}^3P_1$
states for two approximations of exact w.f.} \label{tab.1}
\begin{tabular}{|l|r|r|r|r|}\hline
   nm  &11&  12& 22&32\\
\hline
($w_{nm}$&-0.320& 0.122& - 0.099& -0.0003\\
$5SHO$ &&&&\\
 $w_{nm}$&-0.319& 0.121& - 0.098& -0.0011\\
$3SHO$ &&&&\\
\hline
\end{tabular}
\end{table}

 Then using (\ref{eq.20a})  the w.f. of $X(3872)$ can be presented with a
good accuracy as

\be \varphi(X(3872))= 0.988~\varphi(2\,{}^3P_1) + 0.153~
\varphi(1\,{}^3P_1)\label{eq.17} \ee

%\maketitle

\section{Radiative decays}

Electric dipole transitions between an initial state (i) $n\,{}^3P_1$ state and
a final (f) state $m\,{}^3S_1$ are defined by the partial width \cite{ref.7},
\cite{ref.24},

\begin{equation}
\Gamma(\,\,i  \stackrel{\mathrm{E1}}{\longrightarrow} \gamma +
f\,\,) = \frac{4}{3}\,\alpha\,
e_{Q}^{2}\,E_\gamma^{3}\,(2J_f+1)\,{\rm S}^{\rm E}_{if}
\,|\mathcal{E}_{if}|^{2}~~,~ \label{eq.18}
\end{equation}
where the statistical factor ${\rm S}^{\rm E}_{if}={\rm S}^{\rm E}_{fi}$ is
\begin{equation}
{\rm S}^{\rm E}_{if} = \max{(l,l^{\prime})} \left\{
          \begin{array}{ccc}
            J & 1 & J^{\prime}  \\
            \l^{\prime} & s & l
            \end{array}\right\}^{2}~~.~\label{eq.19}
\end{equation}
For the transitions between the $n\,{}^3P_J$ and
$m\,{}^3S_1~(m\,{}^3D_1)$ states with the same spin $S=1$, the
coefficient  $S_{if}^{E}=\frac{1}{9}(\frac{1}{18})$.

To calculate m.e. $\mathcal{E}_{if}$ we use RSH  $H_0$
\cite{ref.22}, which is simplified in case of heavy quarkonia when
one can neglect a string and self-energy corrections, arriving at
a simple form (widely used in relativistic potential models with
the constituent quark masses in the kinetic term
\cite{ref.25},\cite{ref.26}):

\begin{equation}
H_0=2\sqrt{\vep^2+m_c^2} + V_{\rm B}(r).\label{eq.20}
\end{equation}

By derivation, in (\ref{eq.20}) the mass of the $c$ quark cannot
be chosen arbitrarily and must be equal to the pole mass of a $c$
quark, $m_c\simeq 1.42$~GeV. The pole mass takes into account
perturbative in $\alpha_s(m_c)$ corrections and corresponds to the
conventional current mass $\bar m_c(\bar m_c)=1.22$~GeV
\cite{ref.27} (here $m_c=1424$~MeV is used).

The potential $V_B(r)$ taken,

\begin{equation}
V_{\rm B}(r)=\sigma r - \frac{4\alpha_{\rm B}(r)}{3 r},
\label{eq.21}
\end{equation}
contains the string tension ($\sigma=0.18$~GeV$^2$), which cannot be considered
as a fitting parameter, because it is  fixed by the slope of the Regge
trajectories for light mesons. In the vector strong coupling $\alpha_B(r)$ the
asymptotic freedom behavior is taken into account with the QCD constant
$\Lambda_B$, which is defined by $\Lambda_{\overline{MS}}$:
$\Lambda_B(n_f=4)=1.4238~\Lambda_{\overline {MS}}(n_f=4)=370$~MeV, the latter
is supposed to be known; in our choice $\Lambda_B(n_f=4)$ corresponds to
$\Lambda_{\overline {MS}}(n_f=4)=261$~MeV. At large distances $\alpha_B(r)$
freezes at the value $\alpha_{\rm crit}=0.60$.

Then for a given multiplet $nl$ the centroid mass $M_{cog}(nl)$ is
equal to the e.v. of the spinless Salpeter equation (SSE):

\begin{equation}
H_0\varphi_{nl}=M_0(nl)\varphi_{nl} . \label{eq.22}
\end{equation}
We have calculated $M_{cog} (nl)$ in two cases: in single-channel
approximation, when $M_{cog} (2P) =3954$ MeV was obtained, and
also taking into account creation of virtual loops $q\bar q$,
which are important for the states above the open charm threshold
and give rise to flattening of confining potential \cite{ref.28};
in the last case $M_{cog}(3P)=4295$~MeV,  $M_{cog} (2P) =3943$~MeV
(which is by 9~MeV smaller than without flattening effect), and
then due to fine structure (FS) splittings the mass
$M(2\,{}^3P_1)= 3934$~MeV is calculated.

For a multiplet $nP$  a spin-orbit $a_{so}(nP)$ and tensor $t(nP)$
splittings are calculated here taking  spin-orbit and tensor
potentials as for one-gluon-exchange interaction, although as
shown in \cite{ref.29}, second order ($\alpha^2_{fs} (\mu)$)
corrections appear to be not small for the $1P$ multiplet; their
contribution can reach $ \la 30\%$ .

\be a_{so} (nP) = \frac{1}{2\omega^2_c} \left\{ \frac43
\alpha_{fs} \lan r^{-3}\ran_{nP}- \sigma \lan r^{-1}\ran_{nP}
\right\} + t (nP), \label{eq.23}\ee \be t(nP) = \frac43
\frac{\alpha_{fs}}{\omega^2_c} \lan r^{-3}
\ran_{nP}.\label{eq.24}\ee

We take here  $\alpha_{fs}=0.37$, which provides precise
description of the fine-structure (FS) splittings for the
$1\,{}^3P_1$ charmonium multiplet, if second order corrections are
taken into account \cite{ref.29}. For the $2P$ multiplet the
masses: $M(2\,{}^3P_2) =3963$ MeV, ~~ $M(2\,{}^3P_1) =3934$ MeV, $
M(2\,{}^3P_0) =3885$ MeV,~~ $M(2\,{}^1P_1)=3943$ MeV are obtained.
Notice, that one cannot exclude that for the states above open
charm threshold the FS splittings may be smaller or totally
screened due to coupling to the $DD^*$ channel, and even the order
of the states with different $J$ may be changed.

Below we use the following mass differences:

$$ M_{cog} (2P) - M_{cog}(1P) = 425~{\rm MeV},~~M_{cog} (3P) - M_{cog}(1P) =
770~{\rm MeV},$$\be ~M_{cog} (3P) - M_{cog}(2P) = 350~{\rm MeV},
\label{eq.25}\ee

In Table 2 the m.e. $\mathcal{E}_{if}$ (in GeV) between $n\,{}^3P_1~(n=1,2)$
and $m{}\,{}^3S_1$ states are given; in some cases, if the value of
$\mathcal{E}_{if}$ is small and results strongly depend on $\alpha_{fs}(\mu)$
used, we give two variants: first, with "normal" FS splittings and in second
case FS effects are totally suppressed.

%\end{document}
\begin{table}
\caption{E1 transition rates. The m.e. $\lan
X(3872)|r|n\,{}^3S_1\ran$ $ (n=1,2)$ includes admixture  from the
$1\,{}^3P_1$ component with $c_1 =0.153$; experimental  data from
\cite{ref.30}-\cite{ref.32}.}
\begin{center}
\begin{tabular}{|ll|c|c|c|c|l|l|}\hline
\multicolumn{2}{|c|}{Transition }& $E_\gamma$ & $S_{if}$&
$\mathcal{E}_{if}$&\multicolumn{3}{c|}{ $\Gamma(i\to f)$ (keV)}\\\hline
\multicolumn{2}{|c|}{ $i\stackrel{E1}{\rightarrow}f$}& (MeV) && (GeV)$^{-1}$
&this paper& BG[13]& experiment\\\hline

$1\,{}^3P_1 (3510)~~$&$ 1\,{}^3S_1 (3097)$& 389& $\frac19$& 1.927 & 315&
&$317\pm 25[30]$\\\hline

$2\,{}^3P_1 (3872)~~$&$  1\,{}^3S_1 (3097)$& 697& $\frac19$& 0.104$^{a)}$& 5.3&
11.0
&\\
&&&& 0.216$^{b)}$& 22.8&&\\\hline

$X(3872) $&$ 1\,{}^3S_1 (3097)$& 697& $\frac19$& 0.396$^{c)}$&76.6& &
\\\hline & &&& $>0.292^{d)}$& $>41.7$&&\\\hline

$2\,{}^3S_1 (3686)~~ $&$ 1\,{}^3P_1 (3510)$& 171& $\frac19$& -2.104 & 31.9&
&$30.6\pm 2.2 [31]$\\\hline

$2\,{}^3P_1 (3872)~~ $&$ 2\,{}^3S_1 (3686)$& 181.5& $\frac19$& 3.02 & 78.6&63.9
&\\\hline

$X(3872) ~~ $&$ 2\,{}^3S_1 (3686)$&181.5& $\frac19$& 2.70 & 62.8& &\\\hline

$1\,{}^3D_1 (3770)~~ $&$ 1\,{}^3P_1 (3510)$& 252.9 & $\frac{1}{18}$& 2.767 &
89&199
&$70\pm 17[32]$\\

&&&&&&&$80\pm 24$[27]\\\hline

$2\,{}^3P_2 (3872)~~ $&$ 1\,{}^3D_1 (3770)$& 97.7& $\frac{1}{18}$& -2.776 &
5.2& 3.7&\\\hline

$X(3872) ~~ $&$ 1\,{}^3D_1 (3770)$&97.7& $\frac{1}{18}$& -2.32 & 3.6& &\\\hline
\hline

\end{tabular}
\end{center}
$^{a)}~\alpha_{fs} =0.37$ in spin-orbit potential.~~~~~~~~~~~~~~~~~~~~~~~~~~~~~~~~~~~~~~~~~~~~~~~~~~~~~~~~~~~~~~~~~~~~~~~~~~~~~\\

$^{b)}$ FS interaction is totally suppressed.~~~~~~~~~~~~~~~~~~~~~~~~~~~~~~~~~~~~~~~~~~~~~~~~~~~~~~~~~~~~~~~~~~~~~~~~~~~\\

$^{c)}$ Both admixture of the $1\,{}^3P_1 $ state and $FS$ splittings with
$\alpha_{fs} =0.37$ are taken into account.\\

$^{d)}$ The lower limit refers to the case when FS interaction is totally
suppressed.~~~~~~~~~~~~~~~~~~~~~

\end{table}
To control an accuracy of our calculations in Table 2 we give also
the partial widths of the dipole transitions:
$1\,{}^3P_1\rightarrow J/\psi\gamma,~~2\,{}^3S_1\rightarrow
\chi_{c1}(3510)$, and $1\,{}^3D_1\rightarrow \chi_{c1}(3510)$, and
in all three decays  good agreement with existing experimental
data  \cite{ref.30}-\cite{ref.32} is obtained  (here the
$2\,{}^3S_1$ and $1\,{}^3D_1$ states are identified with
$\psi^{\prime}$ and $\psi^{\prime\prime}$). In our calculations
the $2S-1D$ mixing in not taken into account.

>From Table 2 one can see that for $X(3872)$ as a purely
$2\,{}^3P_1$ state, the transition rate $\Gamma_1\equiv
\Gamma(2\,{}^3P_1\rightarrow J/\psi\gamma)$ strongly depends on FS
potential used.  For this transition the m.e.
$\mathcal{E}_{21}=0.10$~GeV$^{-1}$ is small, if in spin-orbit
potential ``normal"  $\alpha_{fs}=0.37$ is used, and
$\Gamma_1=5.3$~keV is also small, being less  than in
\cite{ref.13}, where a larger $\mathcal{E}_{21}=0.15$~GeV$^{-1}$
and $\Gamma_1=11$~keV were calculated.

In the case when FS interaction is neglected, or suppressed, then
m.e. $\mathcal{E}_{21}=0.216$~GeV$^{-1}$  is larger (in
\cite{ref.13} $\mathcal{E}_{21}\sim 0.276$~GeV$^{-1}$), and the
partial width $\Gamma_1=22.8$~keV is 4 times larger. In  both
cases  discussed a contribution from the $1\,{}^3P_1$ component
was  also neglected and the ratio $R$ is large, $R>4$. Notice,
that in \cite{ref.33}, using an analogy with the radiative decays
of $\chi_{bJ}(2P)$ in bottomonium, a smaller value of this ratio,
$R=1.64\pm 0.25$, was predicted.

However, if admixture from the $1\,{}^3P_1$ state, as in
(\ref{eq.17}), is taken into account, then the transition rate
$\Gamma_1$ increases, independently of a strength of FS
interaction used. With $c_1=0.153$ and suppressed FS interaction
we obtain the lower limit for $\Gamma_1$,

 \be
        \Gamma_1(X(3872)\rightarrow J/\psi\gamma) \geq 41.7~{\rm keV}.  \label{eq.26}
 \ee
This transition rate reaches a larger value, $\Gamma=76.6$~keV, if in FS
potential the same $\alpha_{fs}=0.37$, as for the $1P$ states, is used. On the
contrary, the partial width $\Gamma_2\equiv \Gamma(X(3872)\rightarrow
\psi^{\prime}\gamma)$ decreases (by $20\%$), owing to admixture $c_1$ in the
w.f. of $X(3872)$ and negative  m.e. $\lan 2\,{}^3S_1|r|1\,{}^3P_1\ran$. As a
whole, the ratio $\rm R$ is becoming smaller and  can change in wide range:

 \be
       0.53\leq \rm R \leq (0.8\pm 0.2)~(th),~ \label{eq.27}
 \ee
where the upper limit refers to the case when FS interaction is
strong and theoretical error comes from a variation of
$\alpha_{fs}$, while the lower limit refers to the case when FS
interaction is suppressed.

These values of $\rm R$ are in agreement with the Belle measurements of the
$X(3872)$ radiative decays where a restriction $\rm R< 2.1$ was observed
\cite{ref.16}. At the same time our limit, $\rm R\leq (0.8\pm 0.2)$, does not
agree with  $\rm R=3.4\pm 1.4$ from the BaBar experiment,  being   also much
smaller than in many-channel calculations \cite{ref.13},\cite{ref.14}, where
the ratio $\rm R\simeq 5$ was obtained.

Thus we conclude that precise measurements of $\rm R$ are of great importance
for understanding the nature of $X(3872)$: firstly, for definition of admixture
of the $1\,{}^3P_1$ component in the w.f. of $X(3872)$ and secondly, for
understanding of FS effects in higher resonances, which lie above open-charm
threshold.

Notice, that the partial width $\Gamma_1 (2\,{}^3P_1(3872)  \to
J/\psi\gamma)$ is not very small even for a pure  $2\,{}^3P_1$
state, if a contribution from spin-orbit interaction is
suppressed, and the partial width increases 4 times:

\be \Gamma_1^{(0)} (2\,{}^3P_1(3872)  \to J/\psi\gamma)=22.8~~
{\rm keV}, \label{eq.28} \ee while  in this case  $\Gamma_2^{(0)}
(2\,{}^3P_1(3872)  \to \psi^{\prime}\gamma)=85.5 ~ {\rm keV}$
increases only by $\sim 8\%$; however, their ratio remains large,

\be
 R_0 (2\,{}^3P_1)= 3.75. \label{eq.29} \ee
The situation changes, if FS interaction is suppressed, but  the
w.f. of $X(3872)$  contains admixture of the $1\,{}^3P_1$ state
(\ref{eq.17}). Then the m.e. $\lan X(3872)|r|J/\psi\gamma\ran
=0.512$~ GeV${-1} $ and $\Gamma_1=128$~keV reaches the maximum
value, but the m.e. $\lan X(3872)|r|\psi^{\prime}\gamma\ran
=2.80$~GeV changes by only $\sim 10\%$  and $\Gamma_2=67.6$~keV;
so that their ratio has a minimal value:

\be R_{min})=0.53.\label{eq.30}\ee

To define FS splittings of the $nP$ multiplets we use following
m..e.

$$
 \lan 2P|r^{-1}|1P \ran =0.134~~{\rm GeV},~~\lan 2P|r^{-3}|1P \ran =0.123~~{\rm GeV}^3,$$
 \be \lan 3P|r^{-1}|1P \ran =0.080~~{\rm GeV},~~\lan 3P|r^{-3}|1P \ran =0.112~~{\rm GeV}^3, \label{eq.31}\ee
$$\lan 3P|r^{-1}|2P \ran =0.133~~{\rm GeV},~~\lan 3P|r^{-3}|2P \ran =0.134~~{\rm
GeV}^3.$$ Then  for $\alpha_{fs} =0.37$ one finds
 \be
 \lan X(3872)|r|J/\psi\gamma\ran =0.104~~{\rm GeV^{-1}},\label{eq.32}\ee
which is two times smaller than in the spin-average case, when   $ \lan
X(3872)|r|J/\psi\gamma\ran =0.216$ GeV$^{-1}$, and  the partial width is large,

\be \Gamma_1 (X(3872) \to J/\psi\gamma) = 76.6 ~~{\rm
keV}.\label{eq.33}\ee The m.e. $ \lan
X(3872)|r|\psi^{\prime}\gamma\ran =2.70~~{\rm GeV^{-1}}$ in this
case, giving  \be\Gamma_2 (X(3872) \to \psi^{\prime}\gamma)
=62.8~~{\rm keV},\label{eq.34}\ee and the ratio $\rm R=0.82$.
However,  if  a stronger $\alpha_{fs}\geq 0.45$ is used, this
ratio may reach a larger value, $\sim 1.0$, so that  $\rm R$ can
vary in the range,

 \be R=0.8 \pm 0.2 ~(th).\label{eq.35}\ee
 Thus, while both spin-orbit splitting  and admixture $c_1$ from the $1\,{}^3P_1$
state are present, then  the partial widths $\Gamma_2$ and
$\Gamma_1$ turn out to be of the same order.

In our calculations above we have disregarded the contribution of
the $\gamma$ emission from the $DD^*$ intermediate state in the
radiative decays of $X(3872)$ into $J/\psi$ or $\psi^{\prime}$. To
estimate this part of the $\gamma$ emission we  refer to
calculations done in  \cite{ref.34}. It was found there that the
channel $DD^*$ contributes to the $J/\psi\gamma$ final state less
than $3.6$~keV and to the $\psi^{\prime}\gamma$ final state less
than $0.01$~keV, i.e. these contributions are smaller as compared
to changes in corresponding partial widths due to variations of
the coupling $\alpha_{fs}$ in the range $0.25-0.45$ (see Table 2).
Therefore in this paper we have disregarded possible effect of
$\gamma$ emission  from $DD^*$ intermediate states.

\maketitle

\section{Conclusions}

We study the exotic charmonium state $X(3872)$ with
$J^{PC}=1^{++}$ in the CC approach, where a coupling to the $DD^*$
channels is determined by the parameter-free string-breaking
mechanism. Due to this coupling the $2\,{}^3P_1$ charmonium state
is shifted down to the $D^0D^*$ threshold and its w.f. acquires
admixture from the $1\,{}^3P_1$ $c\bar c$ state. Such mixing of
the $2\,{}^3P_1$ and $1\,{}^3P_1$ states is not large,
corresponding to the mixing angle $\theta=8.8^\circ$.

Owing  to this admixture the transition rate $\Gamma_1(X(3872)\rightarrow
J/\psi\gamma)$ increases several times and reaches the value in the range
$45-80$~keV. At the same time the transition rate $\Gamma_2(X(3872)\rightarrow
\psi^{\prime}\gamma)$ decreases by $\sim 15\%$. As a result their ratio has
following features:

\begin{enumerate}
\item The ratio $\rm R=0.53$, if spin-orbit interaction is totally suppressed.

\item The ratio $\rm R=0.82$, if spin-orbit interaction is defined
by the FS coupling, $\alpha_{fs}\sim 0.37$ and can reach the
larger value $\sim 1.1$ for a larger $\alpha_{fs}$.
\end{enumerate}
The partial width of $X(3872)\rightarrow
\psi^{\prime\prime}\gamma$ appears to be small, $ \sim 4$~keV.

 Our calculations support the Belle result that $\rm
R(exp.) < 2.1$, while a larger number, $\rm R=3.4\pm 1.4$ puts an additional
restrictions on the value of admixture $c_1$ in the w.f. of $X(3872)$.

Acknowledgements.\\
 The authors  are grateful to  Yu.S.Kalashnikova and A.V.Nefediev for useful discussions and suggestions.

The financial support of the Dynasty Foundation to V.D.O. is gratefully
acknowledged.

\end{document}